# Fast Molecular Compression by a Hyperthermal Collision Gives Bond-Selective Mechanochemistry

Lukas Krumbein[†,1], Kelvin Anggara[†,1], Martina Stella[2], Tomasz Michnowicz[1], Hannah Ochner[1], Sabine Abb[1], Gordon Rinke[1], André Portz[3], Michael Dürr[3], Uta Schlickum[1,4], Andrew Baldwin[5], Andrea Floris[6], Klaus Kern[1,7], Stephan Rauschenbach[1,5]*

[1]*Max-Planck-Institut für Festkörperforschung, Heisenbergstrasse 1, DE-70569 Stuttgart, Germany.*

[2]*Department of Materials, Royal School of Mines, Imperial College London, Exhibition Road, London, SW7 2A2, United Kingdom.*

[3]*Institut für Angewandte Physik, Justus-Liebig-Universität Giessen, Heinrich-Buff-Ring 16, DE-35392 Giessen, Germany.*

[4]*Institut für Angewandte Physik, Technische Universität Braunschweig, Mendelssohnstrasse 2, DE-38106 Braunschweig, Germany.*

[5]*Chemistry Research Laboratory, Department of Chemistry, University of Oxford, 12 Mansfield Road, Oxford, OX1 3TA, United Kingdom.*

[6]*School of Chemistry, University of Lincoln, Brayford Pool, LN6 7TS, Lincoln, United Kingdom.*

[7]*Institut de Physique, École Polytechnique Fédérale de Lausanne, Laussane, CH-1015, Switzerland.*

*Corresponding Author: stephan.rauschenbach@chem.ox.ac.uk

[†]equal contributions

## ABSTRACT

Using electrospray ion beam deposition, we collide the complex molecule Reichardt's Dye ($C_{41}H_{30}NO^+$) at low, hyperthermal translational energy (2 - 50 eV) with a Cu(100) surface and image the outcome at single-molecule level by scanning tunneling microscopy. We observe bond-selective reaction induced by the translational kinetic energy. The collision impulse compresses the molecule and bends specific bonds, prompting them to react selectively. This dynamics drives the system to seek thermally inaccessible reactive pathways, since the compression timescale (sub-ps) is much shorter than the thermalization timescale (ns), thereby yielding reaction products that are unobtainable thermally.

## MAIN TEXT

Chemistry is concerned with the manipulation of bonds between atoms with the goal to use chemical reactions to form desired substances. Achieving this goal requires an understanding of how selectivity emerges. Chemical reactions that occur following molecule-surface collisions are, technologically and fundamentally, important in fields as diverse as heterogeneous catalysis [1–9], epitaxial material fabrication [10–12], biomolecular analysis [13–17], and astrochemistry [10]. Bond-selective reactions in a molecule-surface collision have been demonstrated by exciting specific vibrational modes of a molecule right before its surface impact [5,7,8]. This approach succeeds due to the sudden energy accumulation in a specific molecular degree-of-freedom triggering a reaction that promptly occurs before the deposited energy spreads to other degrees-of-freedom not involved in the reaction, i.e. before thermalization [18,19].

Imparting energy into the molecular center-of-mass motion (i.e. translation) towards the surface offers an alternative means to induce surface reactions, which is attractive because translational energy of a molecular ion is readily achieved by acceleration in an electric field. However, simply accelerating molecules towards the surface has been deemed unsuitable for obtaining bond-selective reactions, because the collision excites soft modes of the molecule [20,21] which are poorly coupled to stiff stretching modes that promote bond-breaking reactions [9,19]. As a result, reactions would happen late after the thermalization occurs which diverges reaction pathways towards non-selective outcomes [10,22–26].

It remains unclear whether excited soft modes in the absence of thermalization could give a selective reaction because previous studies have only detected the reaction products in the gas-phase [10,22–26] whereby excess energy from the collision cannot be removed from the molecular product and subsequently causes further reactions. We avoid this issue by using the surface to remove excess energy from the collision products. This is readily achieved by carrying out the experiment at lower, near-threshold energies, which necessitates the detection of the adsorbed collision outcome *on surface*.

Here we show that the large excitation of soft modes in a molecule surface collision, i.e. extensive compression of a molecule, occurring at timescales faster than thermalization, leads to a selective, non-thermal reaction path. We have used scanning tunneling microscopy (STM) to detect the surface-bound products from collisions between a large molecular ion and a metal surface, carried out at low hyperthermal energies (2 – 50 eV) . In our experiment, singly-protonated Reichardt's Dye (**RD**, $C_{41}H_{30}NO^+$) was collided at normal incident angle with a Cu(100) surface held at room temperature using electrospray ion beam deposition (ES-IBD) [14,27,28]. The collision outcome, examined by STM at 11 K (see Supplementary Figs S1), revealed reactive pathways that selectively cleaved a single C-N bond in RD. Energy-dependent experiments and *ab-initio* molecular dynamics (MD) calculations revealed orientation-dependent dynamics that selectively bend specific C-N bonds in the molecule, ignoring the minimum energy path.

Our work gives insight into the emergence of bond-selective mechanochemical reactions in molecule-surface collisions. Collision-induced mechanochemistry [29,30] promises to be generally applicable to large molecules such as peptides or even proteins [13–17], providing a new tool to perform non-thermal on-surface synthesis of novel molecular materials.

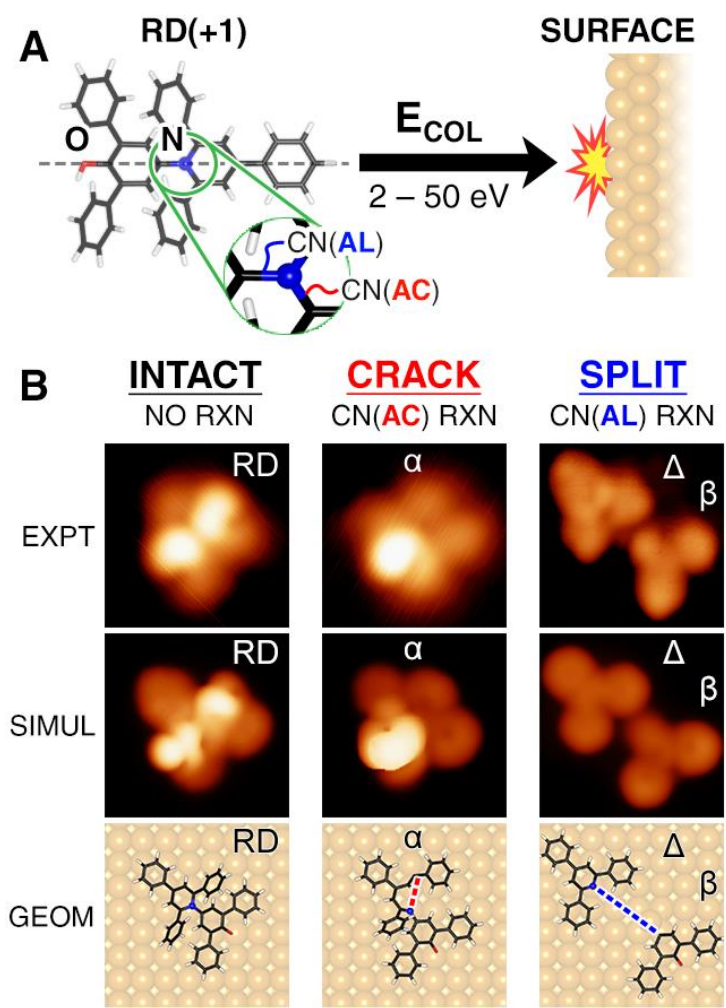

**Fig. 1. | Hyperthermal collision of Reichardt's Dye (RD) on Cu(100) surface**. **A.** Schematics of the experiment, showing a beam of singly-protonated RD(+1) aimed along the surface normal to the Cu-surface at room temperature. Two types of C-N bond are labeled as C-N(AC) and C-N(AL) based on their orientation against the N-O axis in RD (grey dashed line). **B.** STM image (EXPT) and simulation (SIMUL) of the three collision outcomes imaged at 11 K. The INTACT pathway gives adsorbed RD. The CRACK pathway breaks a CN(AC)-bond in the parent RD to give one α-fragment, while the SPLIT pathway breaks the CN(AL)-bond to give one β- and one Δ-fragment. Computed geometries (GEOM) show broken CN(AC)-bond (red dashed line) in the α-fragment, and broken CN(AL)-bond (blue dashed line) between β- and Δ-fragment.

## EXPERIMENT

Figure 1 shows three outcomes obtained from the collision between an RD ion and a Cu(100) surface: one non-reactive, and two reactive outcomes, as revealed by STM imaging and simulation. These are the outcomes of specific collision dynamics that are accessed by aiming the ion beam at normal incidence angle to the surface, as shown in Fig 1A. These three pathways were obtained by analyzing the majority (~80%) of species on the surface. The remaining ~20% of the adsorbate were excluded from our analysis because they cannot be clearly classified (see Supplementary Information). These excluded features can be due to atypical conformations of the identified species at defects, step edges, or in clusters, or they may indicate additional pathways beyond the three discussed here.

The non-reactive pathway, INTACT, was found to yield an adsorbed intact RD with its two phenyl rings oriented vertical from the surface as shown in Fig 1B. We expect the proton that was attached to the O-atom in the RD ion to undergo spontaneous dehydrogenation on surface at room temperature given the low computed barrier of 0.08 eV.

The reactive pathways were found to be bond-selective to the C-N bonds, breaking either one out of the two types of C-N bonds in RD (see Fig 1A): the C-N bond pointing at an angle 'across' the N-O axis of the molecule, termed C-N(AC); or the C-N bond pointing 'along' the N-O axis, termed C-N(AL). In the CRACK pathway, a single C-N(AC) was broken to give a dissociated RD fragment (α-fragment) as shown in Fig 1B. In the SPLIT pathway, the C-N(AL) bond was broken to give two products as shown in Fig 1B: a diphenylphenoxy (β-fragment), and a triphenylpyridine (Δ-fragment). The observation of CRACK and SPLIT pathways thus establishes the existence of bond-selective pathways due to molecule-surface collision at hyperthermal translational energy.

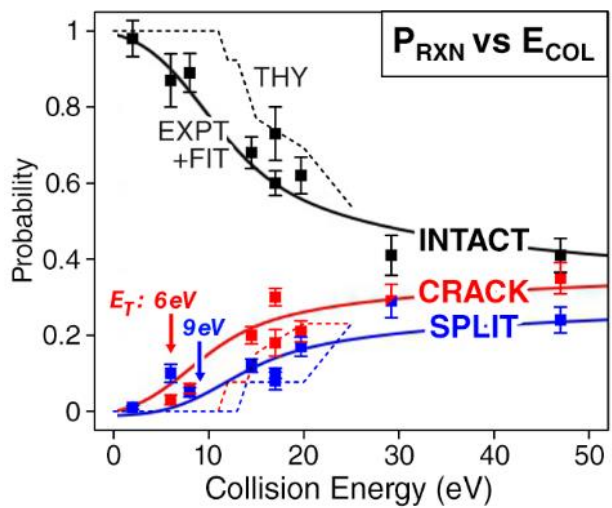

**Fig. 2.** | **Evidence of reactive collision by incident translational energy.** Probabilities to observe INTACT (black), CRACK (red), and SPLIT (blue) outcomes measured against the collision energy. The data points (square) were fitted with a reaction probability model (solid line, eq. 1) to give threshold energy ($E_T$) for each pathways (see Methods for details). Dashed line shows probability obtained from *ab-initio* MD calculations. Error bars show standard deviation.

To gain insight into the dynamics of these bond-selective pathways, we measured their respective reaction probabilities against the kinetic energy of the molecular beam. We varied the collision energy by decelerating the ions approaching the surface, which shifted the energy distribution of the ion beam without changing its width (see Methods and Supplementary Fig S3). Since the ion beam is aimed normal to the surface, the collision energy is the kinetic energy, which corresponds to the molecular translation along the surface normal. The probability of each pathway was obtained by counting the number of species present on surface (see Supplementary Information for details of the analysis, Fig. S2). The result of this energy-dependent measurement was fitted to a model, inspired by the Sudden Vector Projection (SVP) description of molecular collisions at the gas-surface interface [19]. We compute the reaction probability ($P_{RXN,X}$) for each pathway (X) by projecting the molecular translation vector to the reaction coordinate vector to estimate how much of the translation energy ($E_{COL}$) is utilized to propel the system towards the transition state (see Supplementary Fig S4 and Methods for details). The model yields the relation:

$$P_X(E_{COL}) = P_{SAT,X} \cos^{-1} \sqrt{\frac{E_{T,X}}{E_{COL}}}$$

(1)

The fitting gave a translational energy threshold ($E_T$ in Eq. (1)) of 6.0 eV for CRACK, and 9.5 eV for SPLIT, as shown in Fig 2. The existence of this threshold, which marks the minimal energy needed for the reaction, thereby evidences the translational energy as the cause of the reaction, ruling out surface-to-molecule charge transfer [10,24,25] as a sole cause of the reaction.

Further dynamical insight is obtained by the observation of the reaction probabilities approaching a limiting value, $P_{SAT,X}$, estimated to be 0.43 for CRACK, and 0.35 for SPLIT at high energies using Eq. 1. Such a saturation of reaction probability below unity has been observed in gas-surface scattering experiment [25,31,32], and in surface-induced dissociation of proteins [16]. From the diatomic-surface scattering studies [31,32], the saturation was understood to be due to a steric effect in which there was a limited range of orientations in the approaching molecule (also known as 'cone of acceptance' – see ref [33]) that upon collision with the surface would lead to a reactive outcome. Such a strong dependence on initial orientation is characteristic of a *direct* reaction in which a *single-collision* event causes a *direct* energy transfer from the molecule translation into the reaction coordinate [1,31,32,34]. Transferring insights from the diatomic-surface scattering studies to the present work, the saturation in reaction probability observed for CRACK and SPLIT is therefore indicative of an orientation-dependent *direct* reactions caused by a *single-collision* event, ruling out a multi-collision event [34,35] via a precursor state [36].

<u>THEORY</u>

We corroborate this experimental inference by simulating the RD/Cu(100) collision by MD calculations, which revealed that CRACK and SPLIT are orientation-dependent *direct* reaction caused by a *single-collision* event (see Fig 3A and 4, Supplementary Movie 1 and 2). The collision was modeled as a positively-charged RD ion approaching a negatively-charged surface that was understood to contain the image charge of the RD ion (see Supplementary Fig S5). By varying (i) the initial translational energy along the surface normal, and (ii) the initial orientation of the incident molecule (see Fig 3A, shown as the surface approaching the molecule), the calculations reproduced the three outcomes observed in the experiment: the non-reactive INTACT pathway, and the bond-selective CRACK and SPLIT pathways, with a threshold of 12 eV for CRACK, and 14 eV for SPLIT. In our limited sampling of 26 collision geometries (see Fig 3A), the calculations did not reveal additional reaction pathways beyond INTACT, CRACK, and SPLIT. Most notably the computed reaction probabilities were found to reproduce the experimental trend, as shown in Fig 2 and Supplementary Fig S6, thereby validating the collision dynamics unveiled by the MD.

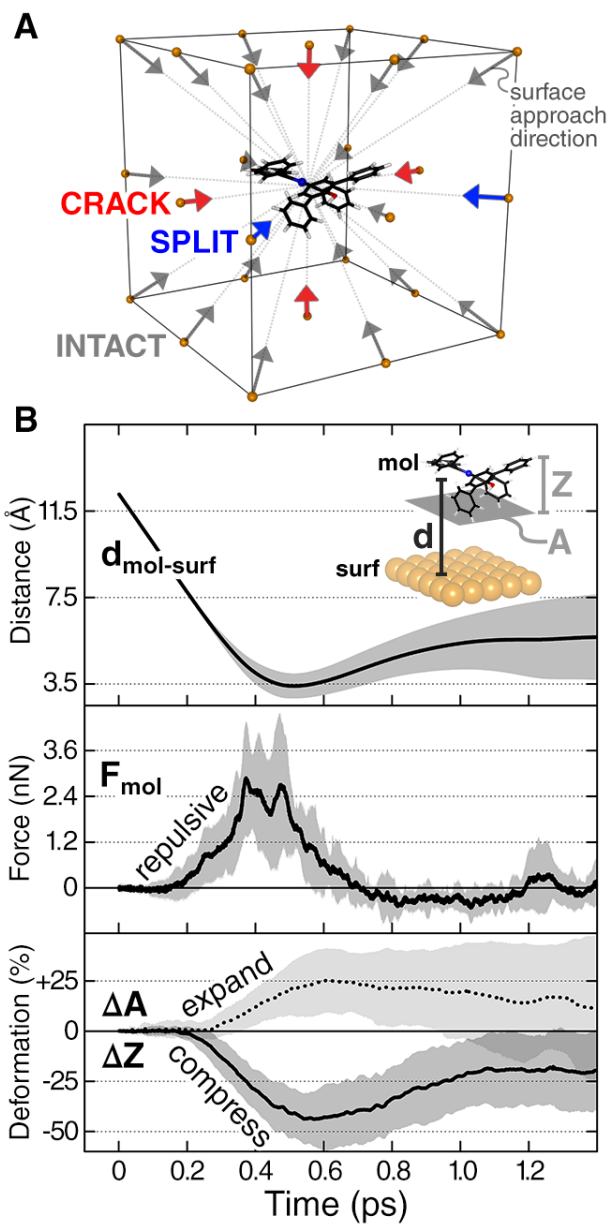

**Fig. 3. | Computed dynamics of RD-Cu(100) collision with varied initial geometries. A.** Schematics of 26 different initial RD geometries simulated for 15 eV collision energy, illustrated here as the surface approaching the molecule (arrows mark the surface normal). Red and blue arrows indicate initial geometries that give CRACK and SPLIT outcomes, respectively; grey arrows for INTACT. **B.** Time-dependent quantities of RD averaged between all trajectories at 15 eV incident energy with different initial geometries: molecule-surface distance ($d_{mol-surf}$), stopping force exerted by the surface on RD ($F_{mol}$), change in RD footprint along the surface plane direction ($ΔA$, dotted line), and the compression of RD along the surface normal ($ΔZ$, solid line). Shaded area is standard deviation of all trajectories.

The collision dynamics for all three pathways in general showed the molecule to be compressed onto the surface due to the mechanical impulse from the molecule-surface impact. As shown in Fig 3B, the stopping force exerted by the surface on the incident molecule was computed to be operative in the nano-Newton regime at sub-picosecond timescale (see also Supplementary Fig S7). This impulse is understood to move the system to a region on the potential energy surface along the compression coordinate. For the reactive CRACK and SPLIT outcome, the system was impulsively propelled to a transition state en route to a product potential well.

The computed dynamics for CRACK and SPLIT show the collision-induced compression precedes the bond-selective dissociation (see Fig 4 and Supplementary Movie 1 and 2). The compression bends a specific C-N bond (i.e. C-N(AC) in CRACK and C-N(AL) in SPLIT) prior to its dissociation (see bottom panels of Fig 4). This C-N bond bending was noted to modify the local geometry of the C-atom from a trigonal planar (sp$^2$ C-atom) to a trigonal pyramidal (sp$^3$ C-atom), in effect converting a p-orbital into a sp$^3$-hybridized C-dangling bond. Upon contact with the surface, this bent C-N bond breaks, and concurrently a new C-Cu bond forms, indicating a reactive event whose transition state is stabilized by the formation of a new bond [37]. In both pathways, the trajectory shows the C-N bond to dissociate in a single attempt, evidencing a *direct* energy transfer from the translational energy to the reaction coordinate.

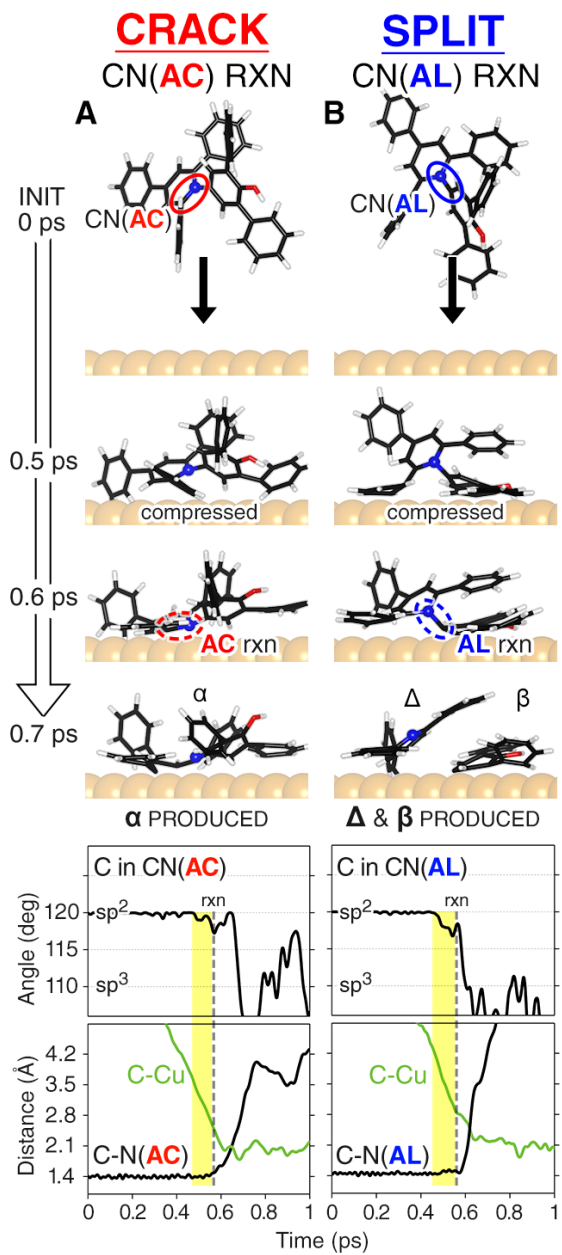

**Fig. 4. | Reactive trajectories for CRACK and SPLIT in RD-Cu(100) collision.** Time-dependent snapshots of the CRACK pathway (**A**) and SPLIT (**B**). Trajectories shown are for the minimum translational energy required to give reactive outcome: 12 eV for CRACK pathway and 14 eV for SPLIT. (**C**) Time-dependent quantities of RD collision: average angle formed between the C-atom in C-N and two of its neighboring atoms, which checks whether the C-atom in the C-N bond is sp²- or sp³-like; C-N distance gives the separation between atoms in the reacting C-N bond (black line); Separation between the C-atom of reacting C-N bond and nearest Cu-atom (green line). The yellow area marks the region where compression from the collision bends the C-N bonds; the dashed line marks the beginning of the C-N rupture.

The preferential breaking of the C-N bond for the bond-selective CRACK and SPLIT pathways was understood to be due to its pronounced reactivity, which we attribute to two factors: (i) the destabilization due to C-N bending, and (ii) the C-N antibonding orbital being the lowest unoccupied molecular orbital (LUMO) of RD (see Supplementary Fig S8). The importance of these two factors were suggested by the correlation between C-N dissociation, and the charge flow from the surface to the LUMO of the compressed RD (i.e. RD with bent C-N bonds) (see Supplementary Fig S9). Similar to mechanism proposed for C-H dissociation of benzene on Cu-surface [38], we expect the orbital mixing caused by the C-N bond-bending increases the propensity of the RD LUMO to hybridize, and form a bond with the surface. The proposed mechanism in which bond-bending alters the electronic structure of a reacting molecule is noted to be topical in mechanochemistry [29,30].

Finally, to address the possibility of the hyperthermal collision giving a chemical reaction via a precursor state of RD, we examined the minimum energy pathway (MEP) of an adsorbed RD. We consider this alternative pathway because an incident RD that had failed to react upon its first collision would have its energy increasingly equipartitioned among all its degrees of freedom while being trapped on surface, allowing the system to search for the MEP to react. The computed barrier along the MEP from an adsorbed RD to give an $\alpha$-fragment was 1.23 eV, and to give a $\beta$- and $\Delta$-fragments, 0.28 eV (see Supplementary Fig S10). The MEP barrier thus predict a reactive outcome dominated by the $\beta$- and $\Delta$-fragments. To test this prediction, we annealed a surface containing adsorbed RD at ~350 K to initiate its thermal reaction (see Supplementary Fig S11). The reaction was found to generate only $\beta$- and $\Delta$-fragments as products, in agreement with the MEP prediction. The hyperthermal collision clearly deviates from the MEP prediction since the α-fragment was the major product, thereby ruling out reactions via precursor state.

In conclusion, we have reported here the first observation of a bond-selective reaction from a hyperthermal collision of a polyatomic ion with a metal surface. The collision gave a mechanical impulse that compresses the incident molecule, causing a specific chemical bond to be activated. From a fundamental standpoint, we expect the collision-induced molecular compression described here to be a general phenomenon across surfaces with similar stiffness; and we expect such mechanism to play a central role in surface-induced dissociation experiments in tandem mass spectroscopy [13–17]. The experimental method described here provides a new general tool to study and apply compressive mechanochemistry for any molecules that can be electrosprayed [27]. Here the compression is operative on the *entire* molecule, instead of locally [39], for subpicoseconds, instead of permanently [30], thereby opening a new avenue to explore *impulsive* mechanochemistry. Our current instrumentation is limited by the finite spread of the translational energy in the incident molecule which introduces uncertainty in the measured threshold energies and in the details of how reaction channels are opened. Furthermore, the control of molecular orientation prior to its surface collision either by using optical [40] or hexapole field [41] should enable selection of reaction pathway. The method described here thus offers new scalable method to non-thermally activate molecules to generate species that are inaccessible from conventional thermochemistry [12,27].

## DATA AVAILABILITY

The authors declare that the data supporting the findings of this study are available from the corresponding author upon request.

## ACKNOWLEDGEMENTS

This work is dedicated to the memory Prof. Alessandro De Vita. We thank Rico Gutzler, John Polanyi, Claire Vallance, Mark Brouard, Christian Reichardt for fruitful discussions. Calculations were performed on the supercomputers at Max-Planck Computing and Data Facility in Garching. The research is funded by the Max-Planck Society. K.A. thanks Alexander von

Humboldt Foundation for financial support. Via our membership of the UK's HEC Materials Chemistry Consortium, which is funded by EPSRC (EP/L000202, EP/R029431), this work used the ARCHER UK National Supercomputing Service (http://www.archer.ac.uk) and the UK Materials and Molecular Modelling Hub for computational resources, MMM Hub, which is partially funded by EPSRC (EP/P020194).

## AUTHOR CONTRIBUTIONS

S.R. designed and supervised the project. S.R., S.A., K.K., U.S. planned experiments. L.K., S.R., T.M., G.R., K.A. performed the experiments. L.K., S.R. H.O., K.A., S.A., A.B. analyzed experimental data. U.S., T.M., A.P., and M.D. performed chemical analysis measurement. K.A., M.S., A.F. performed the DFT calculations.

## COMPETING INTERESTS

The authors declare no competing financial interests.

SUPPLEMENTARY INFORMATION

# Fast Molecular Compression by a Hyperthermal Collision Gives Bond-Selective Mechanochemistry

Lukas Krumbein[†,1], Kelvin Anggara[†,1], Martina Stella[2], Tomasz Michnowicz[1], Hannah Ochner[1], Sabine Abb[1], Gordon Rinke[1], André Portz[3], Michael Dürr[3], Uta Schlickum[1,4], Andrew Baldwin[5], Andrea Floris[6], Klaus Kern[1,7], Stephan Rauschenbach[1,5]*

[1]*Max-Planck-Institut für Festkörperforschung, Heisenbergstrasse 1, DE-70569 Stuttgart, Germany.*

[2]*Department of Materials, Royal School of Mines, Imperial College London, Exhibition Road, London, SW7 2A2, United Kingdom.*

[3]*Institut für Angewandte Physik, Justus-Liebig-Universität Giessen, Heinrich-Buff-Ring 16, DE-35392 Giessen, Germany.*

[4]*Institut für Angewandte Physik, Technische Universität Braunschweig, Mendelssohnstrasse 2, DE-38106 Braunschweig, Germany.*

[5]*Chemistry Research Laboratory, Department of Chemistry, University of Oxford, 12 Mansfield Road, Oxford, OX1 3TA, United Kingdom.*

[6]*School of Chemistry, University of Lincoln, Brayford Pool, LN6 7TS, Lincoln, United Kingdom.*

[7]*Institut de Physique, École Polytechnique Fédérale de Lausanne, Laussane, CH-1015, Switzerland.*

*Correspondence to: stephan.rauschenbach@chem.ox.ac.uk

[†]equal contributions

## **METHODS**

Experiment

A home-built Electrospray Ion-beam Deposition (ES-IBD) apparatus, described in detail in elsewhere [1], was used to collide the Reichardt's Dye (RD) (ref [2]) ion onto the Cu-surface. A dilute solution (~$10^{-4}$ M) of RD (Sigma-Aldrich, 90%) in 3:1 ethanol:water was electrosprayed in positive-mode to generate a beam of monoprotonated $[RD+H]^{1+}$ in the gas-phase. The species in the ion beam (m/z = 552) was mass-selected before its beam energy ($E_{BEAM}$) was characterized by measuring the ion-beam current impinging on an electrode as a function of retarding voltage applied to a grid placed in front of that electrode.

For deposition, the ion-beam was transmitted to an ultra-high vacuum (UHV) chamber (P $\approx 10^{-10}$ mbar) and aimed at normal incident angle onto a clean single-crystal Cu(100) surface held at room temperature. A retarding potential ($V_{DECEL}$) was applied on the Cu-surface to decelerate the incoming ions which controlled the molecule-surface collision energy ($E_{COL} = E_{BEAM} - eV_{DECEL}$, $e$ is elementary charge). Prior to the deposition, the Cu(100) surface was cleaned by repeated cycles of Ar-ion sputtering with 1 keV beam energy and thermal annealing at ~600 K in the UHV environment. After the deposition, the surface was brought to a low-temperature scanning tunneling microscope (Omicron Fermi SPM), where the surface was scanned at 11 K.

Data Analysis: Event Probability

To determine the probability of various outcomes from the molecule-surface collision at a specific $E_{COL}$, we have considered both the adsorbed and scattered collision outcomes. For the latter, we have estimated that scattering of molecules to the gas-phase following their surface collision is not significant due to the good agreement between the total deposited charge and the observed coverage, even at 50 eV collision energy (see Fig S1). Our observation of high sticking probability for large polyatomic ions at 50 eV collision energy is consistent with that reported in the literature [3].

To analyze the adsorbed collision outcome at each $E_{\mathrm{COL}}$, we identify and quantify the adsorbed species with the criterion that the species must adsorb on a flat terrace (see Fig S2 for an example). We classify the adsorbed species into five categories: the intact RD molecule, the alpha ($\alpha$), the beta ($\beta$), the delta ($\Delta$), and the unidentified species (see Fig 1 and Table S1). The unidentified species includes molecules that appear unstable in the STM imaging, and those that were found only once on the surface, such as molecules adsorbed in many different positions next to other molecules in a cluster, or next to surface impurities or defects. The exclusion of small population of species adsorbed on the step-edges in our analysis introduces a small systematic error at every $E_{\mathrm{COL}}$, but should not affect the conclusions drawn from relative comparisons between $E_{\mathrm{COL}}$.

Our observation of intact RD establishes the INTACT pathway, while the alpha species establishes the CRACK pathway, and the beta and the delta species establish the SPLIT pathway. At each $E_{\mathrm{COL}}$, we determine the event probability associated with a pathway ($P_{\mathrm{PATHWAY}}$) by the relative coverage of its product species (i.e. $N_{\mathrm{RD}}$ for INTACT, $N_\alpha$ for CRACK, and $N_\beta$ and $N_\Delta$ for SPLIT):

$$P_{\mathrm{INTACT}}(E_{\mathrm{COL}}) = \frac{N_{\mathrm{RD}}(E_{\mathrm{COL}})}{N_{\mathrm{TOTAL}}(E_{\mathrm{COL}})}$$

$$P_{\mathrm{CRACK}}(E_{\mathrm{COL}}) = \frac{N_\alpha(E_{\mathrm{COL}})}{N_{\mathrm{TOTAL}}(E_{\mathrm{COL}})}$$

$$P_{\mathrm{SPLIT}}(E_{\mathrm{COL}}) = \frac{0.5 \cdot \left(N_\beta(E_{\mathrm{COL}}) + N_\Delta(E_{\mathrm{COL}})\right)}{N_{\mathrm{TOTAL}}(E_{\mathrm{COL}})}$$

$$N_{\mathrm{TOTAL}}(E_{\mathrm{COL}}) = N_{\mathrm{RD}}(E_{\mathrm{COL}}) + N_\alpha(E_{\mathrm{COL}}) + 0.5 \cdot \left(N_\beta(E_{\mathrm{COL}}) + N_\Delta(E_{\mathrm{COL}})\right)$$

where we have included a factor 0.5 for $N_\beta + N_\Delta$ since a single $\beta$-species or $\Delta$-species represent a halved RD molecule, which we found on average in roughly equal quantities ($N_\beta : N_\Delta \approx 1 : 0.8$). We estimate the error associated with each pathway by its standard error ($P_{\text{PATHWAY}}$ / $N_{\text{TOTAL}}^{1/2}$), and thus to ensure statistical significance, we have made ~300 observations (i.e. $N_{\text{TOT}}$) at every $E_{\text{COL}}$.

We have judged that the unidentified species, accounting for ~20% of the species on terrace, cannot serve as evidence to establish nor rule out additional reaction pathways, due to their *non-reproducible* STM appearance preventing their identification. However these minor species does not influence the conclusions presented in the main text, which focuses on the major reactive pathway taken when a molecule collides hyperthermally with a surface. These minor unidentified species, if they originate from the RD/Cu(100) collision, may represent an upper bound of additional error for the absolute outcome probability in every pathway.

| Energy | From STM image counting | | | | | Total outcomes sampled |
| | INTACT | CRACK | SPLIT | | Unidentified | |
| (eV) | $N_{\text{RD}}$ | $N_\alpha$ | $N_\Delta$ | $N_\beta$ | $N_?$ | $N_{\text{TOT}}$ |
| --- | --- | --- | --- | --- | --- | --- |
| 2.0 | 394 | 2 | 3 | 6 | 32 | **400.5** |
| 6.0 | 95 | 3 | 12 | 10 | 57 | **109.0** |
| 8.0 | 251 | 17 | 9 | 18 | 49 | **281.5** |
| 14.5 | 186 | 55 | 24 | 43 | 87 | **274.5** |
| 17.0 | 223 | 112 | 35 | 41 | 160 | **373.0** |
| 17.0 | 84 | 21 | 7 | 12 | 24 | **114.5** |
| 19.7 | 121 | 40 | 35 | 31 | 41 | **194.0** |
| 29.2 | 33 | 23 | 19 | 27 | 48 | **79.0** |
| 47.0 | 54 | 46 | 23 | 39 | 52 | **131.0** |

**Table S1. Number of adsorbed species identified on the surface.** We obtained the relative coverage by identifying and quantifying the species adsorbed on the terrace of the Cu(100) surface. The energy given in the table is the most probable energy of the ion, exemplified by the peak of the energy profile in Fig S3.

| Energy (eV) | Outcome probability | | |
|---|---|---|---|
| | INTACT | CRACK | SPLIT |
| 2.0 | $0.98 \pm 0.05$ | $0.00 \pm 0.00$ | $0.01 \pm 0.00$ |
| 6.0 | $0.87 \pm 0.08$ | $0.03 \pm 0.00$ | $0.10 \pm 0.01$ |
| 8.0 | $0.89 \pm 0.05$ | $0.06 \pm 0.00$ | $0.05 \pm 0.00$ |
| 14.5 | $0.68 \pm 0.04$ | $0.20 \pm 0.01$ | $0.12 \pm 0.01$ |
| 17.0 | $0.60 \pm 0.03$ | $0.30 \pm 0.02$ | $0.10 \pm 0.01$ |
| 17.0 | $0.73 \pm 0.07$ | $0.18 \pm 0.02$ | $0.08 \pm 0.01$ |
| 19.7 | $0.62 \pm 0.04$ | $0.21 \pm 0.01$ | $0.17 \pm 0.01$ |
| 29.2 | $0.42 \pm 0.05$ | $0.29 \pm 0.03$ | $0.29 \pm 0.03$ |
| 47.0 | $0.41 \pm 0.04$ | $0.35 \pm 0.03$ | $0.24 \pm 0.02$ |

**Table S2. Outcome probability as function of collision energy.** The computed probability from the data given in Table S1. The energy refers to the most probable energy of the ion (i.e. the peak of Fig S3).

Data Analysis: Threshold Energies

To fit the experimental data that we inferred to be due to an impulsive collision dynamics, we constructed a mathematical model inspired by the Sudden Vector Projection method that has been applied to model impulsive dynamics in the gas phase and on surfaces [4]. For every reactive pathway, i.e. CRACK and SPLIT, we have postulated:

(1) Unique, non-overlapping reaction coordinate (RC) oriented at a specific axis from the molecule, as illustrated in Supplementary Fig S4.

(2) Each RC-axis possesses a unique threshold energy ($E_T$) for which the reaction would occur if energy supplied along this coordinate is larger than $E_T$.

In the context of a surface approaching the molecule as shown in Supplementary Fig S4, the kinetic energy supplied along a specific RC ($E_{RC}$) was determined by projecting the surface-approach-velocity vector onto the direction vector of the RC, which leads to:

$$E_{RC} = E_{COL} \cos^2 \varphi$$

where $E_{COL}$ is the kinetic energy of the molecule approaching the surface at normal incident angle, and $\varphi$ is the angle between the surface-approach-velocity vector and the RC-vector. We limit the domain of $\varphi$ to $[-\pi/2, \pi/2]$, thereby only considering approach angles that point towards the hemisphere whose axis is the RC-vector (illustrated in Supplementary Fig S4). This allows premise (2) to be expressed as:

$$p_{RXN}(E_{COL}, \varphi) = \begin{cases} 1 & \text{for } E_{RC} \geq E_T \\ 0 & \text{for } E_{RC} < E_T \end{cases}$$

For gaseous molecules impacting on the surface at random orientation as in the experiment, we assume an equiprobable distribution of $\varphi$, expressed mathematically as:

$$f(\varphi) = \frac{1}{\pi}$$

The cumulative probability at a given $E_{COL}$ can therefore be written as:

$$P_{RXN}(E_{COL}) = \int_{-\pi/2}^{\pi/2} f(\varphi) \, p_{RXN}(E_{COL}, \varphi) \, d\varphi$$

Since $p_{RXN}$ is unity only for a range of $\varphi$ that gives $E_{RC} \geq E_T$, we demarcate the maximum range of $\varphi$ by a constant $\varphi_{max}$. As a result, the expression simplifies to:

$$P_{RXN}(E_{COL}) = \frac{1}{\pi} \int_{-\varphi_{max}}^{\varphi_{max}} d\varphi \ = \frac{2\varphi_{max}}{\pi}$$

where:

$$\varphi_{max}(E_{COL}) = \cos^{-1} \sqrt{\frac{E_T}{E_{COL}}}$$

which gives the cumulative probability for a specific RC at a given $E_{COL}$ as:

$$P_{RXN}(E_{COL}) = \frac{2}{\pi} \cos^{-1} \sqrt{\frac{E_T}{E_{COL}}}$$

The expression above only applies to a single RC associated with a single collision outcome. To account for multiple collision outcomes, we have imposed a normalization condition where the sum of probability for all pathways, both reactive and non-reactive, at a specific $E_{COL}$ to be equal to 1, expressed as:

$$P_{INTACT} + P_{CRACK} + P_{SPLIT} = 1$$

We enforce this condition by scaling the event probability by a factor $P_{SAT,X}$ that is unique for each pathway, and independent from $E_{COL}$, thereby yielding the final expression for pathway X (X could be INTACT, CRACK or SPLIT):

$$P_X(E_{COL}) = P_{SAT,X} \, \cos^{-1} \sqrt{\frac{E_{T,X}}{E_{COL}}}$$

... (1)

where $P_{SAT,X}$ is the saturation reaction probability observed at high energy.

To account for the spread in collision energy as in the experiment, we perform a convolution integral in which we replace the exact $E_{COL}$ by a Gaussian distribution centered at $\bar{E}_{COL}$ with a standard deviation of $\Delta E$, yielding the expression of pathway X:

$$P_{RXN,X}(\bar{E}_{COL}) = \int_0^\infty \frac{1}{\sqrt{2\pi}\,\Delta E} \exp\left(-\frac{(E_{COL} - \bar{E}_{COL})^2}{2\Delta E^2}\right) P_{SAT,X} \cos^{-1} \sqrt{\frac{E_{T,X}}{E_{COL}}} \, dE_{COL}$$

... (2)

The equations above were numerically fitted to the experimental results and to the ab-initio calculation results, both yielding good agreement as shown in Supplementary Fig S6. Fitting of Eq (2) to the experiment yielded: $E_{T,CRACK}$ = 5.98 eV, $E_{T,SPLIT}$ = 9.48 eV, $P_{SAT,CRACK}$ = 0.44, $P_{SAT,SPLIT}$ = 0.35; while fitting Eq (1) to the ab-initio MD results yielded: $E_{T,CRACK}$ = 11.22 eV, $E_{T,SPLIT}$ = 14.01 eV, $P_{SAT,CRACK}$ = 0.38, $P_{SAT,SPLIT}$ = 0.39. Here the discrepancy between the threshold energies from experiment and *ab-initio* results can be accounted for by the contribution of Coulomb energy gained due to a molecular ion approaching its image charge missing in the abscissa of Fig 2 and Supplementary Fig S6 (in the EXPT+FIT panel). We estimated the energy to be ~3 eV to be added to the kinetic energy for a point charge approaching its image charge from infinity to its closest-approach distance with the surface (i.e. ~2.5 Å above the image plane). This correction modified the experimental $E_T$ to 9 eV for CRACK, and 12 eV for SPLIT. Furthermore, the threshold energies from ab-initio calculations could be offset by 0.5 - 3.0 eV due to errors in estimating electrostatic interactions with the surface caused by the limited supercell size [5] used in the MD calculations.

Theory: Structure Relaxations, STM Simulations, and Minimum Energy Paths

Density Functional Theory (DFT), STM and Nudged Elastic Band (NEB) [6,7] calculations were performed with the plane wave-pseudopotential package Quantum ESPRESSO [8], using Ultrasoft pseudopotentials [9] with a wave function (charge) kinetic energy cutoff of 408 eV (4080 eV) and a GGA-PBE [10] exchange-correlation functional. The rVV10 [11] self-consistent van der Waals interaction was included. The Brillouin-zone was sampled with the $\mathbf{k} = \Gamma$ point, with a Methfessel-Paxton [12] smearing of 0.27 eV. The Cu(100) surface was modeled with a periodically repeated slab of three layers, with a vacuum gap between the adsorbed molecule and the bottom layer of the slab replica of ~10 Å. Only forces on molecule atoms and surface atoms belonging to the first two layers were allowed to relax, up to 0.026 eV/Å.

STM simulations were performed by calculating the integrated local density of states (ILDOS) within the Tersoff-Hamann method [13]. STM simulations of the intact molecule, α-fragment, Δ-fragment, and β-fragment decorated with a Cu-adatom were obtained at a constant current using the LEV00 package v3.4 (see ref [14]).

NEB calculations are shown in Supplementary Fig S10. In the upper panel we present the Minimum Energy Path (MEP) corresponding to the CRACK reaction of the intact molecule leading to the α-fragment (energy barrier 1.23 eV). In the lower panel, the MEP for the SPLIT reaction to give a β- and Δ-fragments, obtaining a much lower barrier, 0.28 eV. These very different barrier values are compatible with the experimental observations of SPLIT products in the thermal reaction, whereas the CRACK product is observed only in hyperthermal reaction. Note also that while the α-fragment is 0.4 eV less stable than the intact molecule, the β- and Δ-fragments are 0.3 eV more stable. In both NEBs, the state in the reaction coordinate $R_c=1$ corresponds to the intact molecule in Figure 1B (left panel); the $R_c=5$ state in Supplementary Fig S10 (upper panel) to the α-fragment in Figure 1B (central panel) while the $R_c=7$ state in Supplementary Fig S10 (lower panel) to the β- and Δ-fragments in Figure 1B (right panel) without adatom attached to the β.

Theory: Molecular Dynamics

Plane-wave DFT calculations as implemented in Vienna Ab-initio Simulation Package (VASP 5.4.4) (ref [15,16]) were used to model the molecule-surface collision. The calculation used the projector augmented wave method [9,17], Perdew-Burke-Ernzerhof (PBE) functional [10] and Grimme's semiempirical correction for van der Waals interaction (DFT-D3) (ref [18]). The Born-Oppenheimer molecular dynamics (MD) and barrier height calculations were performed at the Γ-point, with a cutoff energy at 400 eV. The Cu(100) surface was modeled as a $(9 \times 8)$ slab with four layers of atoms, where the lowest layer was frozen; and a vacuum gap of 25 Å. The projected-Density-of-States (pDOS) calculations employed a Gaussian smearing with $\sigma = 0.25$ eV. The molecular models and the charge density were visualized using the VESTA software [19].

The MD calculations were performed to simulate the collision dynamics between RD and the Cu(100) surface on the ground potential energy surface. The initial state was a hydrogenated RD with its center-of-mass placed 12 Å above the first layer of the Cu-slab. Each atom in the molecule was given an identical velocity along the surface normal towards the Cu-slab, which sets the center-of-mass translational energy to be the experimental collision energy ($E_{COL}$). Additionally all atoms in the molecule and in the Cu-slab were initialized with random velocities sampled from a Boltzmann distribution at 298

K. The MD simulations were performed with a time-step of 0.5 fs as a microcanonical system that conserved the number of atoms (N), the volume of the supercell (V), and the total energy of the system (E).

The barrier height on the ground potential energy surface for the dehydrogenation of a singly hydrogenated RD on surface was obtained by climbing-image NEB calculations [6,7]. The calculations were performed with at least 5 images between the initial and the final state. The band was relaxed until the forces orthogonal to the reaction coordinate were lower than 0.02 eV/Å.

Theory: Gas-Phase Calculations

The structure of gaseous RD monocation was computed using ORCA [20], employing PBE functional [10], and Grimme's DFT-D3 Van der Waals correction [18]. The calculation was performed using ma-def2-SVP basis sets [21,22], and with the auxiliary basis sets chosen automatically [23].

## SUPPLEMENTARY FIGURES

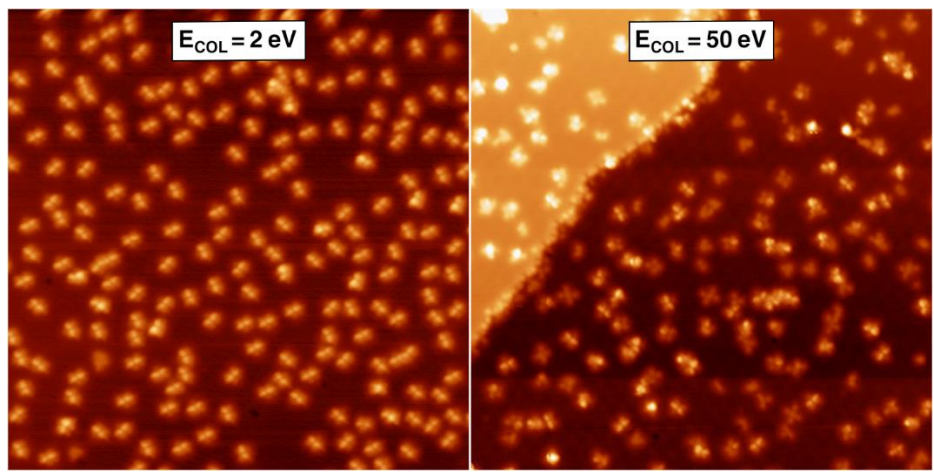

**Supplementary Fig S1. Large-scale image of RD-decorated Cu-surface.** Representative large-scale images ($50 \times 50$ nm$^2$) of RD-decorated surface at two different collision energies ($E_{COL}$). In both cases, the coverage measured from the integrated ion flux was 40 pAh, which corresponds to ~180 molecules per $50 \times 50$ nm$^2$. The difference in surface coverage between these two extreme cases was found to be below ~20%, which can be attributed to anisotropic beam-profile on the surface.

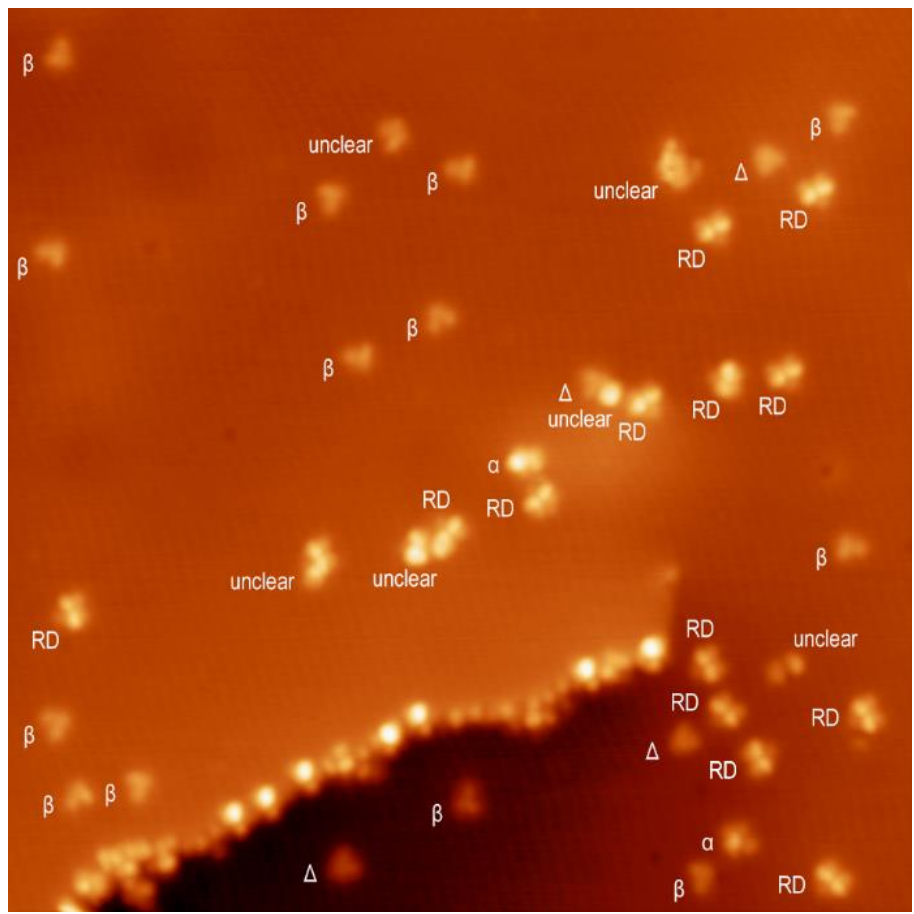

**Supplementary Fig S2. Overview of species present on surface.** An example of image ($50 \times 50$ nm$^2$) analyzed to identify and determine the number of species present on surface after the RD/Cu(100) collision experiment. The collision energy used in this specific example is 15 eV, and in this specific image, we have identified on the surface terrace 13 intact RD molecule, 2 α-species, 13 β-species, 4 Δ-species, and 6 unidentified species. The naming of the species follows the definition in Fig 1, and the definition of the unidentified species follows that outlined in the Data Analysis: Event Probability.

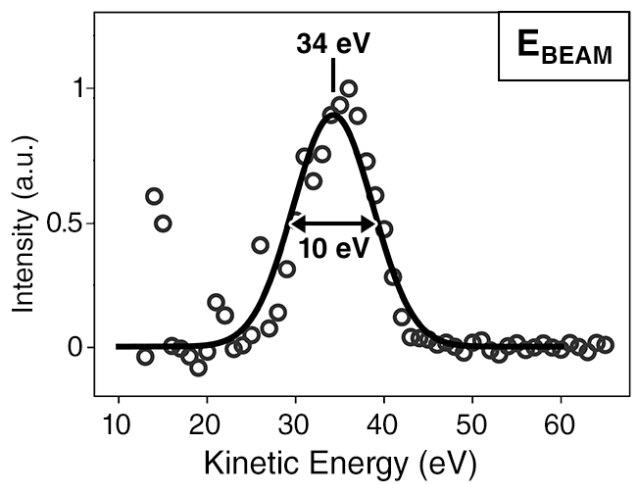

**Supplementary Fig S3. Translational kinetic energy distribution of the molecular ion beam.** A typical energy profile of the RD ion beam. Gaussian fit (solid line) onto the data gives a peak of 34 eV, and an FWHM of 10 eV respectively.

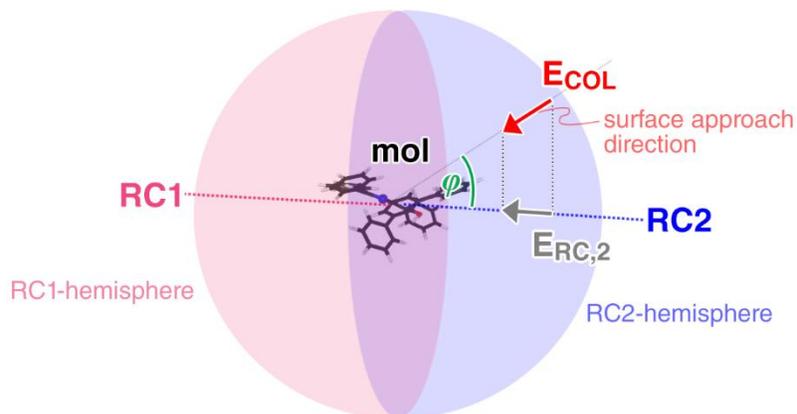

**Supplementary Fig S4. Illustration of collision model.** The outcome of collision is determined by the magnitude of the surface approach vector projected into the reaction-coordinate (RC) – here exemplified by RC1 (in pink) and RC2 (in blue), both shown with their corresponding hemisphere. An example of collision event is defined by an approach vector with a kinetic energy ($E_{COL}$), which forms an angle $\varphi$ (shown in green) with the RC-vector in one of the hemispheres. The projection of the approach vector to the RC-vector was used to compute the kinetic energy that flows along the RC ($E_{RC}$).

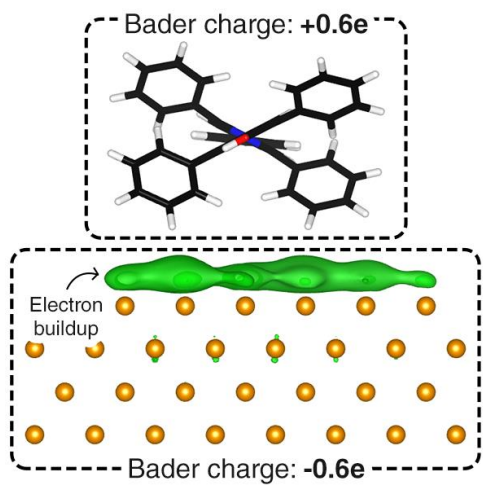

**Supplementary Fig S5. Ground electronic structure of RD/Cu(100) system.** The charge in the molecule located 7 Å above the surface was computed to be +0.6 e, indicating a cationic species, while the charge in the Cu-slab was computed to be –0.6 e. The excess electron in the Cu-slab was computed to mainly localize above the first layer of the surface (green density, isosurface: 0.0002 $e$Å$^{-3}$), as shown by the charge difference density ($\rho_{mol+surf} - \rho_{mol} - \rho_{surf}$) in the Cu-slab. Such a configuration is indicative of charge polarization of a conductor by a nearby point charge.

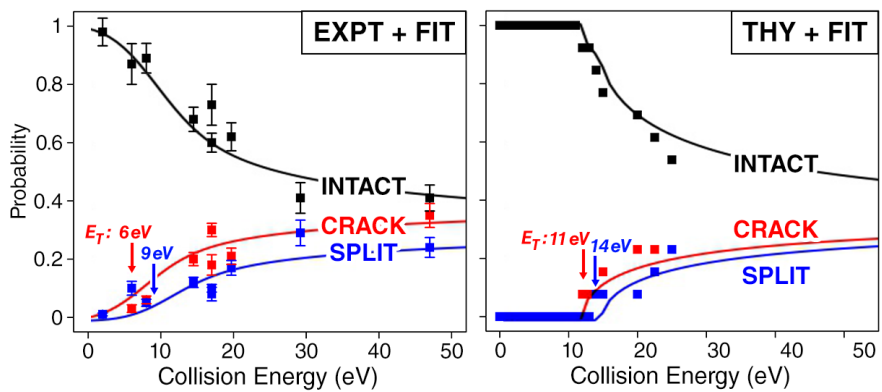

**Supplementary Fig S6. Fitting of collision model into the experiment and ab-initio results.** The fit to experimental data used the energy-profile-broadened version of the fitting equation (see Eq (2) in Methods), while the fitting to ab-initio molecular dynamics results used the non-broadened version of the fitting equation (see Eq (1) in Methods). The fit was performed by enforcing a normalization condition whereby the sum of probabilities from three outcome (CRACK, SPLIT, INTACT) must equal to unity at all collision energies.

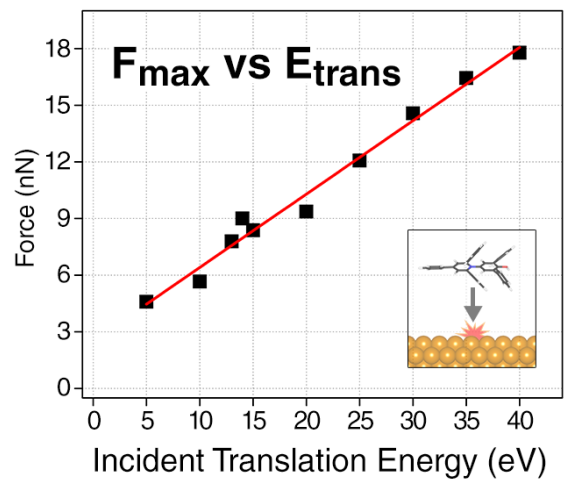

**Supplementary Fig S7. Dependence of maximum stopping force on collision energy.** The peak force obtained from the MD calculations was found to be linear with respect to the collision energy of the molecule to the surface, as shown by the linear fit (red line). The forces shown here are only for a single orientation of RD with its NO-axis parallel to the surface plane (see inset).

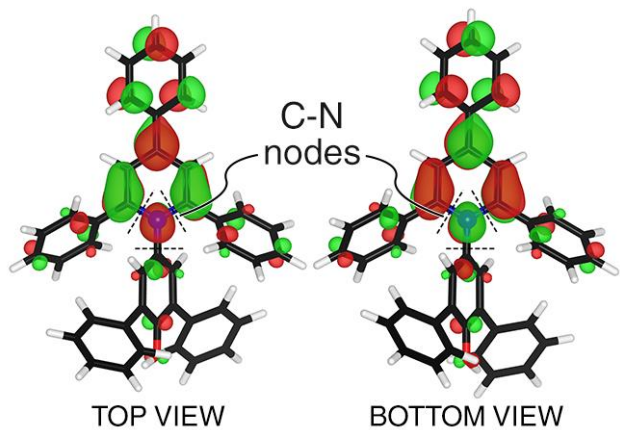

**Supplementary Fig S8. Computed LUMO for a singly-protonated RD in gas-phase.** (A) The LUMO was computed to be a π* antibonding orbital around the C-N bonds (isosurface: 0.03 $e\text{Å}^{-3}$). This is evidenced by the presence of nodal plane (marked by dashed lines) between C- and N-atom in both C-N(AC) and C-N(AL).

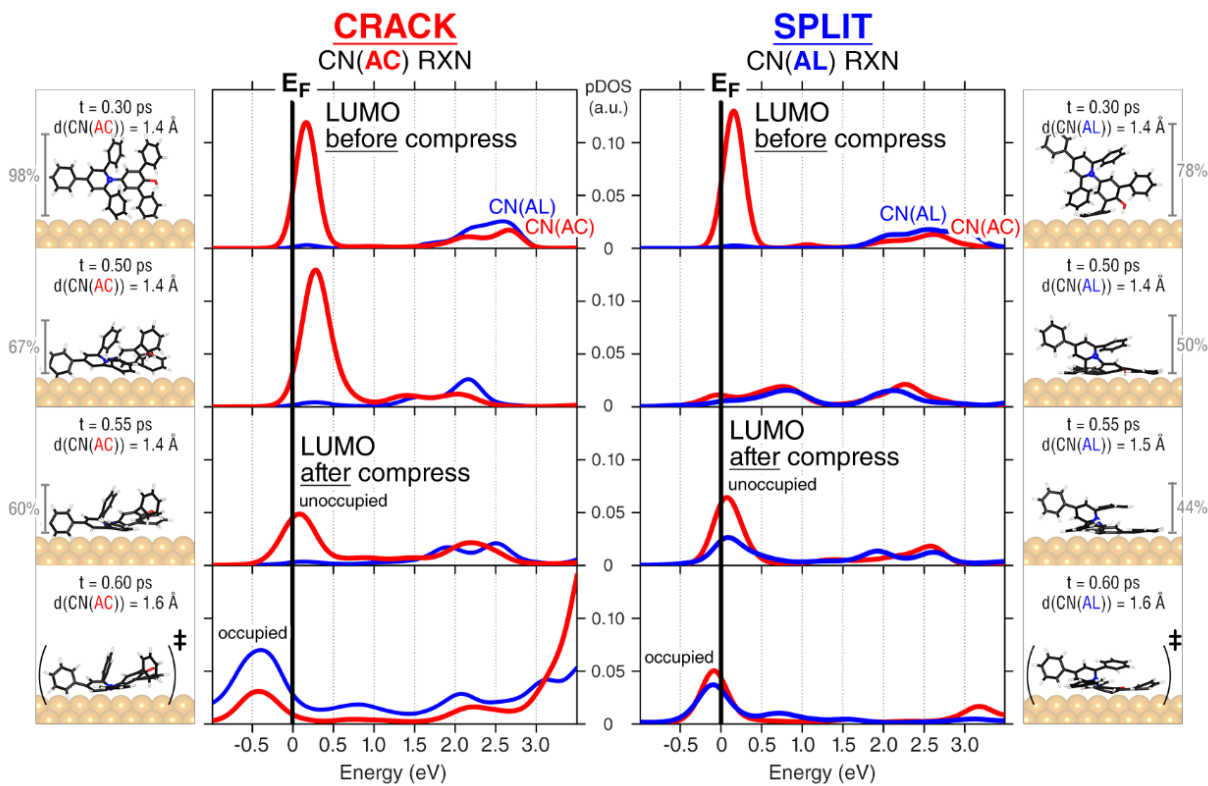

**Supplementary Fig S9. Effect of mechanical compression on RD electronic structure.** The projected density of state (pDOS) shown for the incident RD molecule at different times along the reactive trajectory shown in Fig 4 of the main text. The pDOS shows (i) the compression direction mixes the LUMO differently; and (ii) the surface-to-LUMO charge flow during the concurrent dissociation of the C-N bond and formation of C-Cu bond to the surface (i.e. going from t = 0.55 ps to 0.60 ps). The red line gives the density between the C-atom and the N-atom in C-N(AC), while the blue line, for C-N(AL).

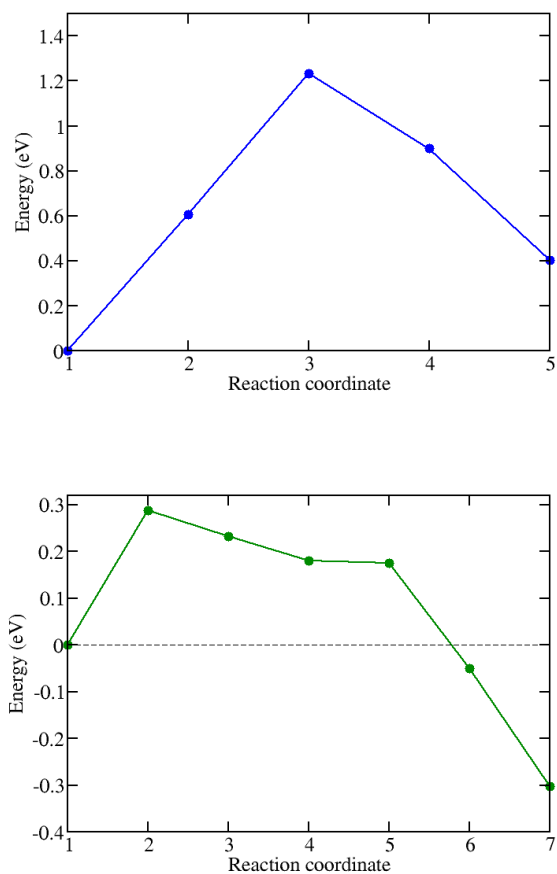

**Supplementary Fig S10. Minimum energy pathways (MEP) and energy barriers for the CRACK and SPLIT.** The upper panel gives the MEP for CRACK, and the lower panel, for SPLIT. Five and seven images (along the reaction coordinate) were used for CRACK and SPLIT, respectively. The barrier is computed to be 1.23 eV for CRACK, and 0.28 eV for SPLIT.

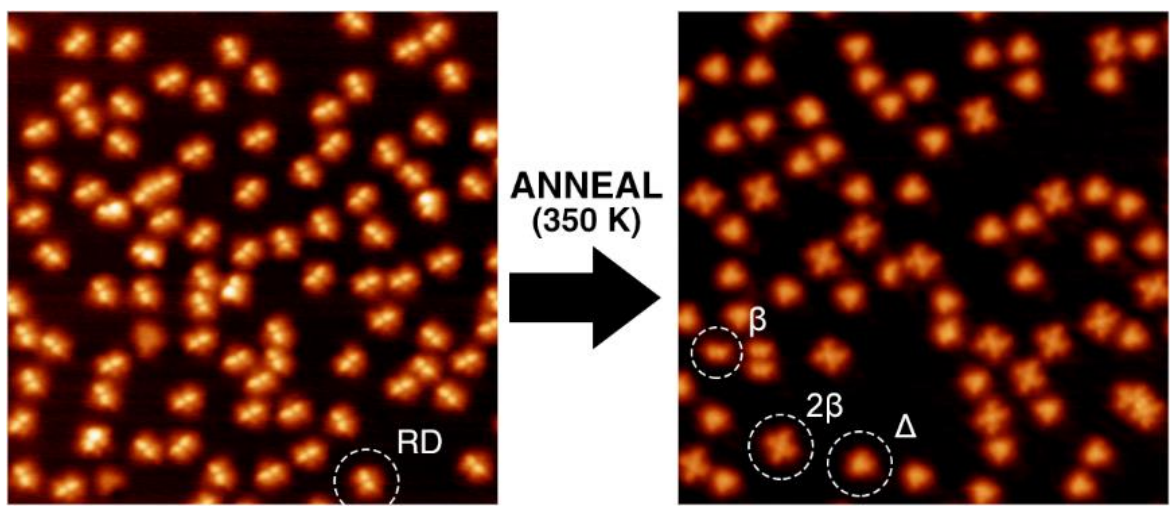

**Supplementary Fig S11. Thermal reaction of adsorbed RD on Cu(100).** STM image at 11 K showing

the surface before and after annealing at ~350 K. At this temperature, adsorbed RD undergoes thermal reaction to exclusively give β- and Δ-species. The initial state is prepared by soft-landing RD ion onto a clean Cu(100) surface held at room temperature. The image sizes are $30 \times 30$ nm$^2$.

**Supplementary Movie 1. Computed dynamics for CRACK pathway.** The trajectory shows C-N(AC) to be broken upon the molecule impacting the surface. The movie was rendered at 30 fps.

**Supplementary Movie 2. Computed dynamics for SPLIT pathway.** The trajectory shows C-N(AL) to be broken upon the molecule impacting the surface. The movie was rendered at 30 fps.

## <u>REFERENCES</u>

VASP, SIESTA, QE and QUICKSTEP.
(https://nms.kcl.ac.uk/lev.kantorovitch/codes/lev00/index.html)